**Malicious aliens and why Caballero's estimates about them should be rejected**

Emory Taylor

Keywords: search for extraterrestrial intelligence, alien life, habitable exoplanets, Kardashev scale, discontinuum physics.

**Abstract**

In a paper published at arXiv.org, Alberto Caballero states a methodology for (1) estimating the probability of an exoplanet alien civilization having malicious intentions toward the human civilization after being messaged by it, and (2) estimating the number of malicious exoplanet alien civilizations in the Milky Way. Caballero states his paper "attempts" to provide these estimates. Caballero's methodology uses questionable hypotheses and excludes important parameters, indicating his methodology and estimates should be rejected.

**Introduction**

Alberto Caballero authored a paper, "Estimating the prevalence of malicious extraterrestrial civilizations", in which he estimates a probability of 5.52E-8% that an extraterrestrial civilization (or fraction of it), on an exoplanet that humanity (i.e., the human civilization) messages, will have malicious intentions towards the human civilization [1]. He also estimates there are 0.22 malicious civilizations in the Milky Way galaxy [1]. To be fair to Caballero, he does state his paper "attempts" to provide these estimates [1]. Caballero claims no serious radio message has been sent to a habitable exoplanet due to a lack of international debate on doing it, and he claims the reason is "a prevailing fear in the scientific community that the message could be picked up by a malicious extraterrestrial civilization [1]. Caballero does not cite a source or indicate these statements are common knowledge.

A Microsoft Bing search on 14 July 2022 for "do scientists fear sending signals to exoplanets" returned results leading to the article "Why These Scientists Fear Contact With Space Aliens" written by Rebecca Boyle [2]. Some astronomers want a beacon sweeping across the Milky Way advertising we are here, and others think the Earth should be a kept secret [2]. Astrophysicist Lucianne Walkowicz thinks it is possible an alien civilization might not have our best interest in mind and thinks "the global scientific community needs to talk about it [2]." Russian billionaire Yuri Milner founded a philanthropic organization called the Breakthrough Initiatives to message aliens [2]. According to Boyle, the big question is whether it is a good idea to find out if there are any aliens by messaging, and quotes Stephen Hawking for whom the answer was "No", indicating the aliens might turn out to be like us [2]. Boyle quoted physicist Mark Buchanan who argued aliens detecting a message from us are likely to be advanced and unfriendly [2]. Boyle points out some scientists think messaging aliens is a good idea and some who once thought it a good idea have changed their minds, and she listed several nonserious messages sent to various targets in space [2]. This led "Several prominent scientists, including Walkowicz, to sign a statement guarding against any future METI efforts until some sort of international consortium could reach agreement [2]."

Caballero has a methodology to "determine how likely it would be that an intelligent civilization living [on] the exoplanet the human civilization messages has malicious intentions and poses a threat to humanity [1]." Caballero states, "[His paper] attempts to provide an estimation of the prevalence of hostile [i.e., malicious] extraterrestrial civilizations through an extrapolation of the probability that we, as the human civilization, would attack or invade an inhabited exoplanet once we become a Type-1 civilization in the Kardashev Scale capable of nearby interstellar travel [1]." Caballero says, "The estimation is based on the world's history of invasions in the last century, the military capabilities of the countries involved, and the global growth rate of energy consumption [1]."

**Questionable hypotheses and estimates based on them**

Caballero used three unstated but implied hypotheses. Caballero's first hypothesis was that the human civilization on earth will behave toward an exoplanet alien civilization in the same way human communities have behaved towards each other, particularly in the 20[th] century [1]. Since there is no historical information on communication between the human civilization and an exoplanet alien civilization or direct (i.e., face-to-face) interaction between these two civilizations, why should Caballero's first hypothesis be accepted? Caballero's second hypothesis was that the Kardashev Scale is correct and applicable [1]. Since the Kardashev Scale is from 1964 [3], and since there have been reevaluations and refutations of the Kardashev Scale [4-7], and since Barrow's scale, which involves mastery over smaller worlds (DNA, molecules, atoms, subatomic particles) [6-7] is an area the human civilization is attempting to master, why should Caballero's second hypothesis be accepted without consideration of Barrow's scale? Caballero's third hypothesis was that the human civilization, after becoming a Type-1 Kardashev Scale civilization in the next 259.5 years, will be capable of interstellar travel such that it can reach an inhabited exoplanet, an untold number of light years away, in order to attack and invade [1]. What are the scientific bases for believing such an undertaking in 259.5 years will be possible without mastering the smaller worlds of DNA, molecules, atoms, and subatomic particles? Caballero did not offer sound scientific reasons for his hypotheses being accepted [1]. Without sound scientific reasons for accepting Caballero's hypotheses, Caballero's estimates, which are based on those questionable hypotheses, should be rejected.

**Limited parameters**

Nikolai Kardashev was an astrophysicist, and he developed the Kardashev Scale [3]. He was concerned with the power a civilization possess and is capable of transmitting, in a coded isotropic radiofrequency, for establishing outer space communication between two civilizations, which are separated by a distance comparable to or greater than the dimensions of the Milky Way galaxy [3]. According to Kardashev, to make an estimate of the rate of information transmission, the transmitter power is the most important parameter [3]. For this reason, Kardashev found it convenient to classify technologically developed civilizations into three types based on energy consumption in erg per second [3]. In a paper published in 2013 in arXiv.org and in Acta Astronautica, "The findings advise caution for proposed interstellar missions, however, as starfaring capability itself might be seen as a threat [by an alien civilization] [8]." Caballero's methodology is based on the human civilization being capable of interstellar travel, but Caballero did **not** consider the prior finding concerning this capability. In 1954, which was approximately ten years before Kardashev published his scale, Einstein speculated that the continuum concept and space and time needed to be discarded and physics should be based on a discontinuum theory [9]. In 2021 the discotninuumphysics.com project was started to follow Einstein's speculations to develop discontinuum physics [10]. By 2022 three discontinuum physics papers were published in the peer-reviewed physics journal Physics Essays [11-13]. In the third of these papers, it was demonstrated that physical reality and nonphysical reality are two inseparable realities [13]. If communications can occur in nonphysical reality, why should the classification of technologically advanced civilizations be based on energy consumption in erg per second? Why should the Kardashev scale continue to be accepted? Caballero has not taken into consideration advancements in physics. Caballero's methodology should be rejected.

Caballer does not address the fact that the human civilization is defensive by nature because humans have the instinct and will to survive, and, for the same reason, it will engage in preparing for possible threats to its way of life and existence. It would be wise to assume the same for an exoplanet alien civilization. Considering these additional facts, the human civilization must consider an exoplanet alien civilization as a possible threat to its way of life and existence and engage in preparing to defend against this threat, at least until such time that it can be determined the exoplanet alien civilization does not pose a threat. It would be wise to assume an exoplanet alien civilization is thinking and doing the same as the human civilization.

For all the human civilization knows the exoplanet alien civilization may have made the decisions that (1) risking its way of life and existence is too big of a risk, and (2) a preemptive strike must be made against any civilization contacting it. The first step in the human civilization's defense is **not** to engage in messaging every exoplanet alien civilization in the Milky Way prior to being a Type-1 civilization that is capable of (1) defending itself, and (2) determining if an exoplanet alien civilization poses a threat prior to contacting it. It would be wise to assume the same of an exoplanet alien civilization. For these reasons, Caballero's methodology should be rejected.

It is common knowledge that (1) the distance from a star where liquid water can exist on a planet is called the habitable zone, or Goldilocks zone, and (2) the human civilization looks for exoplanets in the habitable zone because liquid water is assumed to be a requirement for life. An exoplanet alien civilization faces the impending doom of extinction events as does the human civilization. An important extinction event is stars die, and when they die the habitable zone is not a safe place to be living. Eventually, both the human civilization and an exoplanet alien civilization must find a new planet to call home. What will happen when the Sun is dying and the only habitable zone exoplanet the human civilization can reach is already occupied? Does the human civilization go prepared to exterminate the occupiers if they refuse to share their planet? It would be wise to assume an exoplanet alien civilization has also asked and answered this question. *In a flight from an extinction event for the purpose of survival, the fleeing civilization must harbor malicious intentions or be willing to succumb to extinction*. Caballero did not include extinction events and the instinct and will to survive in his methodology for (1) estimating the probability of an exoplanet alien civilization having malicious intentions toward the human civilization after being messaged by it, and (2) estimating the number of malicious exoplanet alien civilizations in the Milky Way. For this reason, Caballero's methodology and estimates should be rejected.

**Conclusion**

Caballero's methodology used at least three questionable hypotheses without stating sound scientific reasons for accepting them. These hypotheses and the estimates based on them should be rejected. In addition, Caballero's methodology, which is partly based on the human civilization being capable of interstellar travel, should be rejected because it leaves out advancements in physics, extinction events, the instinct and will to survive, and a past finding published in arXiv.org that advised caution for interstellar missions as the capability of interstellar travel might be seen as a threat by an alien civilization.